\documentclass[a4paper, amsfonts, amssymb, amsmath, reprint, showkeys, nofootinbib, twoside]{revtex4-1}
\usepackage[english]{babel}
\usepackage[utf8]{inputenc}
\usepackage[colorinlistoftodos, color=green!40, prependcaption]{todonotes}
\usepackage{amsthm}
\usepackage{mathtools}
\usepackage{physics}
\usepackage{xcolor}
\usepackage{graphicx}
\usepackage[left=23mm,right=13mm,top=35mm,columnsep=15pt]{geometry} 
\usepackage{adjustbox}
\usepackage{placeins}
\usepackage[T1]{fontenc}
\usepackage{lipsum}
\usepackage{csquotes}
\usepackage[pdftex, pdftitle={Article}, pdfauthor={Author}]{hyperref} 
\begin{document}
\title{De Sitter and Minkowski Solutions in 4+1 Brane-World Models}

\author{Josef Seitz}
    \email[Email address: ]{josef.seitz@physik.uni-muenchen.de}
    \affiliation{Arnold-Sommerfeld-Center, Ludwig-Maximilians-Universit\"at,
Theresienstra\ss{}e 37, 80333 M\"unchen, Germany}


\begin{abstract}
We comment on approaches to the Cosmological Constant Problem in 4+1 dimensions, with the Standard Model fields confined to a 3-brane. We show that requiring maximal 4D symmetry only on this 3-brane (and not on all other 3-branes that foliate the 4+1-dimensional space) leads to a wide class of de Sitter solutions on the brane, with the Hubble constant sourced by the brane tension and a bulk scalar potential. For Poincaré symmetry on the brane, fine-tuning is always necessary. The de Sitter solutions are generated at a special dynamical point. Small deviations from this point are shown to grow exponentially with the number of e-folds.
\end{abstract}


\maketitle

\section{Introduction and Outline} \label{sec:outline}
The Cosmological Constant (CC) problem is usually formulated as an issue of fine-tuning: how can the effective CC be as small as $10^{-122}$ (in Planck units) if the expected Standard Model contributions to the vacuum energy are at least $\sim 10^{-60}$? It is, however, well known that a zero CC can be insensitive to the vacuum energy density of the Standard Model fields if the latter live on a 3-Brane embedded into a 4+1 dimensional bulk spacetime, with a single scalar field propagating in the bulk \cite{Arkani-Hamed:2000hpr}\cite{Kachru:2000xs}. That this solution has a bulk singularity is not necessarily bad, as the finite proper distance enforces an effective 4D propagation of gravitons at low energies \cite{Randall:1999vf}. However, it has been shown that this singularity cannot be shielded by a horizon without violating the positive energy condition \cite{Cline:2001yt}. Recently, \cite{Lacombe:2022cbq} have demonstrated that not only a conformal coupling of the bulk scalar (as has been used by \cite{Arkani-Hamed:2000hpr}\cite{Kachru:2000xs}), but also a scalar with Cuscuton \cite{Afshordi:2006ad} action has the ability to render the cosmological constant zero. It has also been shown that no other $k-$essence model \cite{Chiba:1999ka}\cite{Armendariz-Picon:2000nqq} \cite{Armendariz-Picon:2000ulo} can accomplish this. The result seems impressive, especially as there is no obvious fine-tuning involved.\\
In this note, we show that, contrary to the case of the conformally coupled scalar, the Cuscuton also admits an inflationary evolution (i.e. a de Sitter-like solution) on the brane. The subsequent discussion essentially amounts to the question in which sense the model of \cite{Lacombe:2022cbq} is fine-tuned: The Hubble constant of this inflation is sourced by the vacuum energy of the brane and a numerical factor, which has been implicitly set to zero in the case of \cite{Lacombe:2022cbq}.
Additionally, we show that there exists a de Sitter solution for a large class of $k-$essence\cite{Chiba:1999ka}\cite{Armendariz-Picon:2000nqq} \cite{Armendariz-Picon:2000ulo} models, with Hubble rate largely independent of the details of the model. The existence of this solution is provided by the special dynamical point $X=0$, at which the action for the models in question becomes non-analytic. We furthermore show that these inflationary solutions are unstable, in the sense that small deviations from $X=0$ grow with $\sim (e^{\# e-folds })^c$ (with $c$ a numerical factor).
By introducing a bulk potential, these solutions can also be tuned to achieve Poincaré symmetry on the brane. The reason why these solutions have not been identified previously is that we require maximal symmetry only on the $y=0$ slice and not on all $y=const$ slices.  
We also comment on possible further investigations. 
\section{The Model} \label{sec:themodel}
We consider a model in the spirit of \cite{Randall:1999vf}\cite{Randall:1999ee} by finding solutions of the Einstein equations with metric:
\begin{equation} \label{eq1}
    ds^2 = a^2(y,\eta)\eta_{\mu \nu}dx^{\mu} dx^{\nu}+dy^2
\end{equation}
in the 'mostly +' convention. For the fifth coordinate $y$, orbifold boundary conditions are imposed, which (practically speaking) amounts to $a(-y,\eta)=a(y,\eta)$. \\
We want to emphasize that \ref{eq1} is, in general, not a good choice of metric to find solutions of the Einstein equations, as it is on one hand too general to respect maximal 4D symmetry, on the other hand, is too restrictive for only spatial symmetry. The proper, more general approach would be to consider \cite{Mukohyama:1999qx}:
\begin{equation}\label{eq2}
    ds^2=-a^2(y,\eta) d\eta ^2+b^2(y,\eta)dx^i dx_i +dy^2
\end{equation}
Nevertheless, for finding either a Minkowski solution ($a(y=0,\eta)=a(y=0)$) or a de Sitter solution ($a(0,\eta)=-\frac{a(0)}{H_0 \eta}$) on the brane, this form of metric is sufficient and easy to calculate with. 
As in \cite{Lacombe:2022cbq}, we consider the bulk action 
\begin{equation}\label{eq3}
    S = \int d^5x \sqrt{-g}(\frac{R}{\kappa_5^2}+P(X,\phi))
\end{equation}
with $X=-\frac{1}{2} \partial_{N}\phi \partial_{M}\phi g^{MN}$ and $\phi$ a scalar field, where we follow the usual convention that five-dimensional indices are denoted by Roman letters. We also have the usual brane action 
\begin{equation}\label{eq4}
    S_{brane} = \int \sqrt{-g_4} V_0 e^{2\kappa_5 \phi}
\end{equation}
i.e. the coupling of the scalar to brane matter fields is only a Weyl transformation artefact in the Jordan frame, satisfying the Weak Equivalence Principle, exactly as in \cite{Arkani-Hamed:2000hpr}\cite{Kachru:2000xs}\cite{Lacombe:2022cbq}. $V_0$ is the vacuum energy on the brane. The key difference to \cite{Lacombe:2022cbq} is now that we allow for a time dependent scalar field, i.e.
\begin{equation}\label{eq5}
    X= \frac{1}{2a^2}\dot{\phi}^2-\frac{1}{2} \phi'^2
\end{equation}
The Israel junction condition\cite{Israel:1966rt} is the same as if we had no time-dependence in the metric, as there are no $\delta-$function-like charges in the $\eta$-direction. The same holds for the scalar field junction condition. Hence we get (as in \cite{Lacombe:2022cbq}):
\begin{align}\label{eq6,7}
    & \frac{a'(0^+)}{a}= -\frac{\kappa_5^2}{6}V_0 e^{2\kappa_5  \phi}\\
    & \phi'(0^+)= \frac{V_0\kappa_5}{P_X} e^{2\kappa_5 \phi}
\end{align}
where we have suppressed the $\eta$-dependence. The resulting four independent Einstein equations are: 
\begin{align}\label{eq8-11}
    &3a'^2+3a''a+(\frac{\dot{a}}{a})^2-2\frac{\ddot{a}}{a}=\kappa_5^2(-p_5 a^2) \\
    &-3a'^2-3a''a+3(\frac{\dot{a}}{a})^2=\kappa_5^2(p_5 a^2+(\rho_5+p_5)\frac{\dot{\phi}^2}{-2X}) \\
    &6(\frac{a'}{a})^2-3\frac{\ddot{a}}{a^3}=\kappa_5^2(-p_5 +(\rho_5+p_5)\frac{\phi'^2}{-2X})\\
    &-3(\dot{\frac{a'}{a}}) = \kappa_5^2 (\rho_5+p_5)\frac{\phi' \dot{\phi}}{-2X}
\end{align}
with $\rho_5=P(X,\phi)-2XP_X(X,\phi)$, $p_5= -P(X,\phi)$; $'$ denotes the derivative with respect to $y$, while $\dot{}$ denotes the derivative with respect to $\eta$.\\ 
If we are only interested in possible solutions at the brane (and not in the whole five-dimensional bulk), we can set $y=0^+$ and insert the junction conditions. After adding two equations, we get:
\begin{align}\label{eq12-15}
    &\frac{1}{a^2} ( 4(\frac{\dot{a}}{a})^2 -2 \frac{\ddot{a}}{a}) = P_X (2X+(\frac{\kappa_5}{P_X}V_0 e^{2\kappa_5 \phi})^2) \\
    & G_{\eta y}= \kappa_5^2 T_{\eta y}=0 \\
    & \frac{\kappa_5^2 e^{4\kappa_5 \phi } V_0^2}{6} - \frac{3}{\kappa_5^2}\frac{\ddot{a}}{a^3} = P + \kappa_5^2 V_0^2 e^{4\kappa_5 \phi}/P_X \\
    & 3a'^2+3a''a+(\frac{\dot{a}}{a})^2-2\frac{\ddot{a}}{a}=\kappa_5^2(-p_5 a^2) 
\end{align}
where we have also used
\begin{equation}\label{eq16}
    \frac{\dot{\phi}^2}{a^2}=2X+\phi'^2    
\end{equation}
We want to emphasize that solutions of the Einstein equation obtained in this way make no statement about solutions in the bulk; in particular it is not possible to say whether maximal symmetry is preserved on the brane.
The second equation of \ref{eq12-15} is trivially fulfilled (this is intuitive, as $G_{\eta y}=R_{\eta y}=R^{M} \,_{\eta M y}$ and $R^{M} \,_{\eta N y}$ can be associated with a closed loop in the $y-\eta$ plane, intersecting the brane) while the last equation just determines the continuation of the solution at the brane to the bulk (notice that it is the only equation containing $a''$, which is not fixed by junction conditions). \\
Then, as expected, we arrive at two modified Friedmann equations in conformal time: \\
\begin{align}\label{eq17,18}
    &\frac{1}{a^2} ( 4(\frac{\dot{a}}{a})^2 -2 \frac{\ddot{a}}{a}) = P_X (2X+(\frac{\kappa_5}{P_X}V_0 e^{2\kappa_5 \phi})^2) \\
    & \frac{\kappa_5^2 e^{4\kappa_5 \phi } V_0^2}{6} - \frac{3}{\kappa_5^2}\frac{\ddot{a}}{a^3} = P + \kappa_5^2 V_0^2 e^{4\kappa_5 \phi}/P_X  
\end{align}
As our predominant interest is in flat and de Sitter solutions, the right-hand side of the first equation should vanish (as for $a\propto 1/\eta$ the left-hand side does). As previously mentioned, there is an important difference between our considerations and the considerations within most of the standard literature on brane-world models \cite{Arkani-Hamed:2000hpr}\cite{Randall:1999vf}\cite{Randall:1999ee}\cite{DeWolfe:1999cp}: in the case of those, the 4D maximal symmetry is present in all 4-dimensional slices of the bulk with constant $y$; here we relax this condition and only check for 4D maximal symmetry at $y=0$, i.e. at the brane. Other slices in the bulk can have lower symmetry. This idea is slightly similar to the models brought forward to explain inflation on the brane-like \cite{Maartens:1999hf}\cite{Copeland:2000hn} (see \cite{Brax:2003fv} \cite{Maartens:2010ar} for reviews on brane-world cosmology), but with a critical difference: The scalar field in our case is free to propagate in the bulk.\\ 
Further considerations depend on the specific model of the scalar field. Consider the case
\begin{equation}\label{eq19}
    P=F(\phi) (-X)^{\alpha}-V
\end{equation}
i.e. a homogeneous $k-$essence field with extra potential. Then the r.h.s. of equation (16) becomes
\begin{equation}\label{eq20}
    2\alpha F(\phi) (-X)^{\alpha} + \frac{\kappa_5^2 V_0^2e^{4\kappa_5 \phi}}{-2\alpha F(\phi)(-X)^{\alpha-1}}
\end{equation}.
For generic $\phi$ and generic functions $F(\phi)$ both terms have to vanish; this is only the case for $X=0$ and $0<\alpha <1$ (but see section \ref{sec:pointX}). In this case, \ref{eq17,18} yields together with $a=-\frac{1}{H_0 \eta}$
\begin{equation}\label{eq21}
    H_0^2= \frac{\kappa_5^4 V_0^2 e^{4\kappa_5 \phi}}{36}+\frac{\kappa_5^2 V(\phi)}{6}
\end{equation}
The inflation on the brane can be driven by both a linear bulk term and a quadratic brane tension term. This relation for the Hubble constant is in spirit similar to the standard result \cite{Csaki:1999jh}\cite{Cline:1999ts} that the expression for the square of the Hubble constant in Randall-Sundrum type models \cite{Randall:1999vf} contains a term quadratic in the brane energy density. The Hubble constant is independent of the kinetic term and its precise shape, which is expected from the point of view of standard inflationary theory.\\
The point $X=0$ is a special point in the dynamics of the scalar, as the action is not analytic there. We now analyze it and the corresponding inflationary solution further.

\section{The point $X=0$} \label{sec:pointX}
From (16) and the junction condition on the scalar field, we get a further constraint on $\alpha$. For $\alpha<\frac{1}{2}$ 
\begin{align}\label{eq22}
\begin{split}
    & \frac{\dot{\phi}^2}{a^2}=2X+\phi'^2 = 2X+(\frac{\kappa_5}{-\alpha (-X)^{\alpha-1} F(\phi)}V_0 e^{2\kappa_5 \phi})^2 \\ 
    & \xrightarrow{X \rightarrow 0^-} 2X < 0 \\
\end{split}
\end{align},
a contradiction. Only $\frac{1}{2} \leq \alpha < 1$ is possible. As we will treat the extremal case $\alpha = \frac{1}{2}$ in detail in IV, we restrict to $\alpha > \frac{1}{2}$ for now. \\
To understand what happens in the limit $X\rightarrow 0$, we can analyze the symplectic form of the scalar field for a point on the brane. The canonical momentum reads
\begin{align}\label{eq23}
    \begin{split}
        \Pi(y=0) = &-\frac{\dot{\phi}}{2a^2} F(\phi)\alpha (-X)^{\alpha-1} \\
        = & - sgn(\dot{\phi}) \frac{\alpha (-X)^{\alpha-1}F(\phi)}{2a} \sqrt{2X+\phi'^2} \\
    \end{split}
\end{align}.
\textit{Close to} $X=0$, this becomes 
\begin{equation}\label{eq24}
    \Pi(y=0) \approx \frac{-\kappa_5 V_0 e^{2\kappa_5 \phi}}{2a} sgn(\dot{\phi})
\end{equation}
and the symplectic form vanishes; there is no additional dynamics on the brane. \textit{At} $X=0$, \ref{eq22} demands that $\dot{\phi}$ vanishes for $\alpha <1$, restoring the dynamics of the scalar on the brane (as there is a $\delta-$function-like dependence on $\dot{\phi}$). It is these special dynamics that drive the inflation; from this perspective, it is also clear why \cite{Arkani-Hamed:2000hpr} did not find any de Sitter solution for the brane: for $\alpha=1$ (and only for $\alpha=1$) and $V(\phi)=0$, which corresponds to their choice of action, $\phi'$ is independent of $X$; \ref{eq17,18} demands $\dot{\phi}=0$, which in turn leads to a finite, non-zero $X$ completely determined by the junction condition. The field is therefore not dynamical, independent of the value of $X$; the behaviour of the solution (Minkowski or de Sitter) is completely specified by $F(\phi)$. The Hubble constant becomes
\begin{equation}\label{eq25}
    H_0^2 = \frac{\kappa_5^4 V_0^2 e^{4\kappa_5 \phi}}{12} (\frac{1}{3}-\frac{1}{F(\phi)})
\end{equation}
$F(\phi)=3$ yields the Minkowski solution on the brane, as expected. This makes the role of fine-tuning in \cite{Arkani-Hamed:2000hpr} very transparent: the only place where it is needed is for $F(\phi)=3$ ($V(\phi)=0$ should not be considered fine-tuning, as it can be justified by a shift symmetry). As we will see in section $IV$, the situation is somewhat similar in the case $\alpha = \frac{1}{2}$.\\

\section{$\alpha = 1/2$ and the role of Fine-Tuning in the Cuscuton Model} \label{sec:Cuscuton}
In the case $\alpha=\frac{1}{2}$, $\phi'(y=0) \propto \sqrt{-X}$ and the canonical momentum on the brane becomes
\begin{equation}\label{eq26}
    \Pi(y=0) = -sgn(\dot{\phi}) \frac{(-X)^{\alpha -1}}{4a} \sqrt{2X+\phi'^2} \propto (-X)^0
\end{equation},
it depends only on $\phi$ (up to a step function of $\dot{\phi}$). In this case, there are no additional local degrees of freedom on the brane as long as $\dot{\phi}\neq 0$ (the reasoning here is similar to an argument in the original introduction of the Cuscuton \cite{Afshordi:2006ad}). This is an immediate consequence of the fact that 
\begin{equation}\label{eq27}
    \frac{\phi'^2}{-2X}=\frac{2\kappa_5^2 V_0^2}{F^2(\phi)}e^{4\kappa_5 \phi} \equiv \frac{1}{1-\delta(\phi)}
\end{equation}
is independent of $X$. $1-\delta(\phi)$ relates a brane property (the tension) to a bulk property ($F(\phi)$). The appearance of $\delta(\phi)$ is unique to the case $\alpha=\frac{1}{2}$; in all other cases the above ratio diverges in the $X\rightarrow 0$ limit.\\
In this limit, the two "Friedmann equations" simplify and again lead to \ref{eq21}.
\subsection{Poincaré invariance and Self-tuning Cuscuton solution}\label{Cuscuton_fine_tuned}

If we want to restore Poincaré invariance on the brane, the two contributions of \ref{eq21} have to cancel out and (if the cancellation is to occur for generic $\phi$) 
\begin{equation}\label{eq28}
    V(\phi)= - \frac{\kappa_5^2 V_0^2 e^{4\kappa_5 \phi}}{6}
\end{equation}
Combining with equation (27) yields the condition
\begin{equation}\label{eq29}
    V(\phi)=-\frac{F^2(\phi)}{12(1-\delta(\phi))}
\end{equation}
As previously mentioned, $1-\delta$ relates brane properties to bulk properties; a Minkowski solution on the brane then requires that two bulk potentials ($F(\phi)$ and $V(\phi)$) must be related by the brane tension. From this point of view, there is no $\delta(\phi)$ that is more natural than any other; the bulk potentials need to be necessarily fine-tuned to accommodate a Minkowski solution. We want to emphasize that the issue of fine-tuning is less severe if \ref{eq28}\ref{eq29} are only required to hold for a specific value of $\phi$, then some mechanism could relax $\phi$ towards this specific value. However, this is not realized in the two known self-tuning mechanisms \cite{Arkani-Hamed:2000hpr}\cite{Lacombe:2022cbq}. (The heuristic explanation is that $\phi$ varies through the bulk and that \ref{eq28}\ref{eq29} therefore have to hold for generic $\phi$. Of course, this is not a valid argument, as in slices with $y \neq 0$ there is no brane tension.) Constructing such a model is left for future work. \\
We recover the more specific self-tuning Cuscuton solution of \cite{Lacombe:2022cbq} by picking $\delta(\phi)=0$: Defining $c_0 \equiv \frac{\kappa_5^2 V_0^2}{6}$ yields the scalar field bulk action 
\begin{equation}\label{eq30}
    P(X,\phi)= - 2\sqrt{-3c_0 e^{4\kappa_5 \phi} X} + c_0 e^{4\kappa_5 \phi} 
\end{equation},
where we have, as in \cite{Lacombe:2022cbq}, chosen the sign for $F(\phi)$ consistent with causality conditions on the Cuscuton field \cite{Afshordi:2006ad}. 

From the analysis of \cite{Lacombe:2022cbq}, we know that in the case of $\delta(\phi)=0$ the Minkowski symmetry is apparent in every constant $y$ slice, which is a result not accessible from our "pure brane" considerations. That the $\delta \neq 0$ solutions have not been found signals that those do not preserve Poincaré symmetry in all constant $y$ slices. Although unusual, it is not inconceivable that this can happen, as $y=0$ corresponds to a boundary of the space and is hence not a generic slice. 
\subsection{de Sitter solutions}\label{Cuscuton de Sitter}

Using the junction condition (27) leads to the Hubble constant 
\begin{equation}\label{eq31}
    H_0^2 = \frac{\kappa_5^2}{6}(V(\phi)+\frac{F^2(\phi)}{12(1-\delta(\phi))})
\end{equation}
This nicely shows the role of fine-tuning involved in the solution of \cite{Lacombe:2022cbq}: their action fulfills the condition $V(\phi)=-\frac{F(\phi)^2}{12}$, which leads to the Hubble constant
\begin{equation}\label{eq32}
    H_0 = \frac{\kappa_5^2 V_0 e^{2\kappa_5 \phi}}{6} \sqrt{\delta(\phi)}
\end{equation}
and tuning $\delta$ to zero again leads to the self-tuning mechanism. We want to emphasize that at the critical point $X=0$, there is no symmetry justification to do this: $\dot{\phi}=0$ is independent of $\delta(\phi)$.

\section{Stability analysis of the de Sitter solution} \label{sec:Stability}
There is extensive literature on perturbation theory in brane-world models (see, for example, section 6 of \cite{Maartens:2010ar} and references therein). Here we just check whether the inflationary solutions found in \ref{sec:themodel} \ref{sec:Cuscuton} are stable against small deviations from the point $X=0$. We consider the Cuscuton solution of \ref{Cuscuton de Sitter} separately. This is by no means a full perturbation analysis, we instead consider a very specific perturbation mode (the one which keeps the 4D theory conformally flat and 3D symmetric). If the solution is unstable with respect to this mode, then it is of course also unstable in the full perturbation analysis. \\

\subsection{Stability of $\frac{1}{2}<\alpha < 1$}
Consider the perturbation
\begin{align}\label{eq33,34}
& \phi = \phi_0 + \kappa_5 \delta \phi(\eta,y=0)\\
& a=-\frac{1+\delta a}{H_0 \eta}    
\end{align}
around an inflationary solution of \ref{eq17,18} with Hubble constant $H_0^2=\frac{\kappa_5^4 V_0^2 e^{4\kappa_5 \phi_0}}{36}$ (setting $V$ to zero for simplicity; the qualitative behaviour is not different for $V\neq 0$). The linearized Einstein equations with $\kappa_5=1$ become 
\begin{align}\label{eq35,36}
    &\eta \dot{\delta a}+\frac{1}{2}\eta^2 \ddot{\delta a} = \frac{9}{\alpha F(\phi_0)} (-\delta X)^{1-\alpha} \\
    & 4 \delta \phi + 2\delta a + \eta \dot{\delta a}-\frac{1}{2} \eta^2 \ddot{\delta a} = - \frac{6}{\alpha} (-\delta X)^{1-\alpha} \frac{1}{F(\phi)}
\end{align} 
where we have taken into account that $\delta X^{\alpha} \ll \delta X^{1-\alpha}$. $\delta X$ is given by the junction condition as
\begin{equation}\label{eq37}
-\delta X=(\pm \alpha \frac{F(\phi)}{6} \eta \dot{\delta \Tilde{\phi}})^{\frac{1}{1-\alpha}}    
\end{equation}
where the sign depends on the sign of $\delta \dot{\phi}$. This results in
\begin{align}\label{eq38,39}
    & \eta \dot{\delta a}+\frac{1}{2} \eta^2 \ddot{\delta a}=\pm \frac{3}{2} \eta \dot{\delta \phi} \\
    & 2\delta \phi + \delta a + \eta \dot{\delta a} = \pm \frac{5}{4} \dot{\delta \phi}
\end{align}
These equations have a constant solution $2 \delta \phi = -a$, which corresponds to a shift of the Hubble constant. The non-constant solution is a polynomial in $\eta$ with one decaying solution (i.e. $\propto \eta^s$, $s>0$) and one solution that grows. It is interesting to note that since the exact power depends on the sign in \ref{eq38,39}, the precise degree of instability depends on whether there is a perturbation with $\delta \phi >0$ or $\delta \phi <0$. In both cases, however, the scale factor deviates strongly from de Sitter after only a finite number of e-folds. 

\subsection{Stability of Cuscuton Inflation} 
For simplicity, we consider the case $\frac{-F^2(\phi)}{12}=V(\phi)$ and non-zero, but constant $\delta$ (compare \ref{Cuscuton de Sitter}). Constant $\delta$ can be assumed without loss of generality, as any variation of it is in a higher order of the perturbation.   
We use again expansion \ref{eq33,34}. The resulting linearized equations are
\begin{align}\label{eq40,41}
    & -4 \eta \dot{\delta a} - 2\eta^2 \ddot{\delta a} = 6\sqrt{2} \frac{\kappa_5}{H_0} \sqrt{-X} \sqrt{\frac{\delta}{1-\delta}} \\
    & 4 \delta \phi + 2 \delta a-\frac{1}{2}\eta^2\ddot{\delta a}+\eta \dot{\delta a} = \frac{\sqrt{2}\kappa_5}{H_0} \sqrt{-X} \sqrt{\frac{\delta}{1-\delta}}
\end{align}
From the junction condition, one gets
\begin{equation}\label{eq42}
    \kappa_5 \sqrt{-2X}=\pm \sqrt{\frac{1-\delta}{\delta}}H_0 \eta \dot{\delta \phi}
\end{equation}
again depending on the sign of the perturbation. The perturbation equations become completely independent of $\delta$. Inserting a polynomial Ansatz of the same power for $\delta a$, $\delta \phi$ leads to a decaying solution ($\delta a \propto (-\eta)^4$ for the positive sign in \ref{eq42} and $\delta a \propto (-\eta)^5$ for the negative sign) and a growing solution ($\delta a \propto (-\eta)^{-4}$ for the positive sign in \ref{eq42} and $\delta a \propto (-\eta)^{-1}$ for the negative sign). Even though the deviation in both cases grows with conformal time (and hence with $e^{\# e-folds}$), there is a difference: one perturbation grows much more slowly than the other.

\section{Summary and Outlook}
We have commented on the role of fine-tuning in the context of solving the CC problem with a bulk scalar in 4+1 dimensions, with the Standard Model fields confined to a 3-brane. In particular, we have shown that there is a dynamically interesting point $X=0$, at which there exists an inflationary solution on the brane for a large class of scalar field actions. The Hubble parameter is largely independent of the details of the model. It contains both contributions from the bulk scalar potential and the brane tension (for example, for a homogeneous $k$-essence field we have just $H_0 \propto V_0 e^{2\kappa_5 \phi}$), and to avoid this solution (meaning setting $H_0$ to zero) requires fine-tuning between contributions from the brane and contributions from the bulk, as expected from \cite{Randall:1999vf}. We highlight this in the example of the recently found\cite{Lacombe:2022cbq} Cuscuton\cite{Afshordi:2006ad} solution that can produce a vanishing CC. The only way to avoid this fine-tuning would be to find a mechanism or UV theory that naturally connects the brane tension with the scalar field bulk action. We have also demonstrated that the found inflationary solutions tend to be unstable with respect to small deviations from $X=0$.\\
So far we have only considered models which incorporate a de Sitter/Minkowski solution for generic field values $\phi$. It might be entirely possible that a model can be justified which has a Minkowski/de Sitter solution only at a single field value $\phi$, as long as the scalar field theory has some mechanism that relaxes it to this value. This could be potentially phenomenologically relevant, as the Hubble constant scales exponentially with $\phi$; large negative values of $\phi$ ($-\mathcal{O}(100)$ in Planck Units) could then suppress the Hubble scale to its small current value. A further stability analysis of this class of solutions would be in order. 



\begin{thebibliography}{4}
\bibitem{Arkani-Hamed:2000hpr}
N.~Arkani-Hamed, S.~Dimopoulos, N.~Kaloper and R.~Sundrum,
Phys. Lett. B \textbf{480} (2000), 193-199
doi:10.1016/S0370-2693(00)00359-2
[arXiv:hep-th/0001197 [hep-th]].

\bibitem{Kachru:2000xs}
S.~Kachru, M.~B.~Schulz and E.~Silverstein,
Phys. Rev. D \textbf{62} (2000), 085003
doi:10.1103/PhysRevD.62.085003
[arXiv:hep-th/0002121 [hep-th]].

\bibitem{Randall:1999vf}
L.~Randall and R.~Sundrum,
Phys. Rev. Lett. \textbf{83} (1999), 4690-4693
doi:10.1103/PhysRevLett.83.4690
[arXiv:hep-th/9906064 [hep-th]].

\bibitem{Cline:2001yt}
J.~M.~Cline and H.~Firouzjahi,
Phys. Rev. D \textbf{65} (2002), 043501
doi:10.1103/PhysRevD.65.043501
[arXiv:hep-th/0107198 [hep-th]].

\bibitem{Lacombe:2022cbq}
O.~Lacombe and S.~Mukohyama,
JCAP \textbf{10} (2022), 014
doi:10.1088/1475-7516/2022/10/014
[arXiv:2203.16322 [hep-th]].

\bibitem{Afshordi:2006ad}
N.~Afshordi, D.~J.~H.~Chung and G.~Geshnizjani,
Phys. Rev. D \textbf{75} (2007), 083513
doi:10.1103/PhysRevD.75.083513
[arXiv:hep-th/0609150 [hep-th]].

\bibitem{Chiba:1999ka}
T.~Chiba, T.~Okabe and M.~Yamaguchi,
Phys. Rev. D \textbf{62} (2000), 023511
doi:10.1103/PhysRevD.62.023511
[arXiv:astro-ph/9912463 [astro-ph]].

\bibitem{Armendariz-Picon:2000nqq}
C.~Armendariz-Picon, V.~F.~Mukhanov and P.~J.~Steinhardt,
Phys. Rev. Lett. \textbf{85} (2000), 4438-4441
doi:10.1103/PhysRevLett.85.4438
[arXiv:astro-ph/0004134 [astro-ph]].

\bibitem{Armendariz-Picon:2000ulo}
C.~Armendariz-Picon, V.~F.~Mukhanov and P.~J.~Steinhardt,
Phys. Rev. D \textbf{63} (2001), 103510
doi:10.1103/PhysRevD.63.103510
[arXiv:astro-ph/0006373 [astro-ph]].

\bibitem{Randall:1999ee}
L.~Randall and R.~Sundrum,
Phys. Rev. Lett. \textbf{83} (1999), 3370-3373
doi:10.1103/PhysRevLett.83.3370
[arXiv:hep-ph/9905221 [hep-ph]].

\bibitem{Mukohyama:1999qx}
S.~Mukohyama,
Phys. Lett. B \textbf{473} (2000), 241-245
doi:10.1016/S0370-2693(99)01505-1
[arXiv:hep-th/9911165 [hep-th]].

\bibitem{Israel:1966rt}
W.~Israel,
Nuovo Cim. B \textbf{44S10} (1966), 1
[erratum: Nuovo Cim. B \textbf{48} (1967), 463]
doi:10.1007/BF02710419

\bibitem{DeWolfe:1999cp}
O.~DeWolfe, D.~Z.~Freedman, S.~S.~Gubser and A.~Karch,
Phys. Rev. D \textbf{62} (2000), 046008
doi:10.1103/PhysRevD.62.046008
[arXiv:hep-th/9909134 [hep-th]].

\bibitem{Maartens:1999hf}
R.~Maartens, D.~Wands, B.~A.~Bassett and I.~Heard,
Phys. Rev. D \textbf{62} (2000), 041301
doi:10.1103/PhysRevD.62.041301
[arXiv:hep-ph/9912464 [hep-ph]].

\bibitem{Copeland:2000hn}
E.~J.~Copeland, A.~R.~Liddle and J.~E.~Lidsey,
Phys. Rev. D \textbf{64} (2001), 023509
doi:10.1103/PhysRevD.64.023509
[arXiv:astro-ph/0006421 [astro-ph]].

\bibitem{Brax:2003fv}
P.~Brax and C.~van de Bruck,
Class. Quant. Grav. \textbf{20} (2003), R201-R232
doi:10.1088/0264-9381/20/9/202
[arXiv:hep-th/0303095 [hep-th]].

\bibitem{Maartens:2010ar}
R.~Maartens and K.~Koyama,
Living Rev. Rel. \textbf{13} (2010), 5
doi:10.12942/lrr-2010-5
[arXiv:1004.3962 [hep-th]].

\bibitem{Csaki:1999jh}
C.~Csaki, M.~Graesser, C.~F.~Kolda and J.~Terning,
Phys. Lett. B \textbf{462} (1999), 34-40
doi:10.1016/S0370-2693(99)00896-5
[arXiv:hep-ph/9906513 [hep-ph]].

\bibitem{Cline:1999ts}
J.~M.~Cline, C.~Grojean and G.~Servant,
Phys. Rev. Lett. \textbf{83} (1999), 4245
doi:10.1103/PhysRevLett.83.4245
[arXiv:hep-ph/9906523 [hep-ph]].




\end{thebibliography}
\end{document}